\documentclass{ws-mpla}
\usepackage{graphicx}
\usepackage[T2A]{fontenc}
\usepackage{epigraph}
\usepackage{hyperref}
\usepackage{url}
\usepackage{color}
\usepackage{soul}
\hypersetup{colorlinks,urlcolor=cyan,citecolor=green,linkcolor=blue,filecolor=black}
\begin{document}

\catchline{}{}{}{}{}

\title{Einstein, tea leaves, meandering rivers, and the origin of supermassive black holes in ekpyrotic universe}
\author{Z.K.Silagadze}

\address{Budker Institute of Nuclear Physics and Novosibirsk State University, 630 090, Novosibirsk, Russia \\  silagadze@inp.nsk.su}

\maketitle

\pub{Received (Day Month Year)}{Revised (Day Month Year)}

 
\begin{abstract}
Central supermassive black holes are found in most massive galaxies. However, their origin is still poorly understood. Observations of quasars show that many supermassive black holes existed less than 700 million years after the Big Bang. To explain the existence of such black holes with masses comparable to the stellar mass of the host galaxy, just ~500 million years after the Big Bang, it is probably necessary to assume that they originated from heavy seeds. In an ekpyrotic universe, a hot Big Bang occurs as a result of the collision of two branes. Quantum fluctuations create ripples on the brane surfaces, leading to spatial variations in the timing of collisions, thereby creating density perturbations that can facilitate the formation of massive black hole seeds. I hypothesize that perhaps Rayleigh–B\'{e}nard-Marangoni type convection in the extra dimension is a more efficient source of macroscopic density perturbations than quantum fluctuations.
\end{abstract}


\maketitle
%
\section{Foreword} 
\epigraph{Нам тайны нераскрытые раскрыть пора, -\\
Лежат без пользы тайны, как в копилке.\\
Мы тайны эти с корнем вырвем у ядра,\\
На волю пустим джинна из бутылки!\\
It’s time to disclose undisclosed secrets, -\\
Secrets lie in vain like in a pig bank.\\
We well uproot these secrets from the nucleus, \\
We will set free the genie from the bottle!     
}{V. Vysotsky, 1964. Translated by A. Sokolov.}

The idea came about while listening to Alexander Dolgov's talk ``Primordial black holes, seeding of cosmic structures, and dark matter and antimatter''  at the Quarks-2024 conference, in which he argued that his and Silk's thirty-year-old idea \cite{Dolgov_1993} that galaxy formation occurs under the influence of supermassive primordial black holes is gaining increasing support. After the talk, I told him about Einstein's tea leaf effect and whether something similar might have helped form seeds in the early universe. "That's a very interesting observation," he replied. I wasn't sure if it was just a polite remark or if he was genuinely interested in the idea, and I had no intention of publishing anything about this curious fantasy.

But after the conference I visited Dubna and in the Vysotsky caf\'{e} I saw on the wall a fragment of his poem “March of Physics Students”, which I use as an epigraph. The metaphor of a genie being let out of a bottle captured my imagination, and I decided to free the genie that had taken up residence in my dreams, even though it was not yet fully formed. In any case, it could not cause any harm, unlike the nuclear genie.

\section{Introduction}
The $\Lambda$CDM concordance model of cosmology proposes that the large-scale structure of the Universe is shaped by gravitational instability due to small initial density inhomogeneities arising from quantum fluctuations amplified by inflation \cite{Dodelson_2021}. 

The $\Lambda$CDM cosmological model is a remarkable achievement: with only six parameters, it describes the evolution of the Universe from the earliest times to the present day in agreement with many precise observations \cite{Turner:2018}. However, as the precision of cosmological observations has increased, a number of problems with this standard paradigm have emerged \cite{Perivolaropoulos:2021}.

Observations with the James Webb Space Telescope show that the Universe younger than 1 billion years and even 500 million years is populated by objects that, according to conventional wisdom, are unlikely to have appeared there in such a short time \cite{Dolgov_2018}. In particular, it is argued that the observed co-evolution of galaxies and supermassive black holes implies the existence of a large population of massive black holes at very early times \cite{Silk_2024}. However, as is well known, the formation of supermassive black holes at large redshifts $z\ge 7$ cannot be explained if the growth of stellar-mass black hole embryos by accretion is limited by Eddington accretion \cite{Woods_2019,Volonteri_2010}. New observational data appear to point to the opposite picture compared to the $\Lambda$CDM cosmological model: supermassive black holes are not created in galactic halos, but rather galaxies form around primordial black hole seeds \cite{Dolgov_2018}.

Primordial black holes are thought to have formed in the early Universe through various processes \cite{Carr:2003,Carr:2023}, and their masses are estimated to be roughly equal to the mass contained in the particle horizon at the redshift of their creation: from the Planck mass for black holes formed in the Planck epoch, to $M_\odot$ for black holes formed during the QCD phase transition, and up to $10^5M_\odot$ \cite{Volonteri_2010,Khlopov:2004}.

The observed exponential increase in the ultraviolet luminosity with redshift is difficult to reconcile with the predictions of the $\Lambda$CDM model without modifying the model by introducing a bump at the $k\sim 1\mathrm{Mpc}^{-1}$ scale in the power spectrum of the primordial density perturbations \cite{Padmanabhan_2023,Tkachev:2023}. In this letter, the hypothetical mechanism generating the macroscopic density perturbations is qualitatively considered within the framework of the ekpyrotic universe scenario.

\section{Tea leaf effect and meandering rivers}
Have you ever wondered why rivers meander? Einstein wondered and gave some explanations \cite{Einstein_1926}\footnote{You can find the English translation here \cite{Yogananda_2000}, and the Russian translation here \url{https://ufn.ru/ru/articles/1956/5/j/}}. Einstein is famous for his thought experiments. However, this time he describes a real experiment that anyone can repeat: ``Imagine a flat-bottomed cup full of tea. At the bottom there are some tea leaves, which stay there because they are rather heavier than the liquid they have displaced. If the liquid is made to rotate by a spoon, the leaves will soon collect in the center of the bottom of the cup'' \cite{Einstein_1926}. Einstein's explanation of this phenomenon, with the addition of some missing details, is as follows.

\begin{figure}[h]
    \centering
    \includegraphics[scale=0.3]{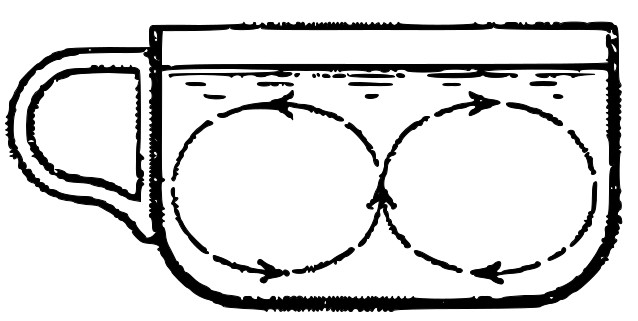}
    \caption{Einstein's illustration of secondary circulation in a cup.}
    \label{fig1}
\end{figure}

When the tea rotates, the centrifugal force pushes the liquid outward, deforming the free surface: the lowest height of the liquid column is in the center of the cup and increases towards the walls of the cup. As a result, a pressure gradient arises that balances the centrifugal force in a rotating system, where the bulk of the liquid is at rest. However, due to friction against the walls of the cup and the bottom, part of the liquid in the boundary layers is carried away by the cup and therefore moves in the rotating system.  The resulting Coriolis forces acting on the elements of the liquid are directed towards the center of the cup and lead to the emergence of secondary circulation, schematically shown in Fig.\ref{fig1} from Einstein's article. As a result, ``the tea leaves are swept into the center by the circular movement and act as proof of its existence'' \cite{Einstein_1926}.

In fact, Einstein's explanation was slightly different. In the boundary layer at the bottom, the liquid rotates on average more slowly than in the bulk.  So in the new rotation system, in which the boundary layer fluid is on average at rest, the centrifugal force is smaller, but the pressure gradient remains the same. Therefore, this smaller centrifugal force can no longer balance the pressure gradient, and the fluid in the boundary layer begins to move toward the center. To support this flow, the liquid near the free surface streams away from the center and descends along the side walls of the cup.

But what does all this have to do with meandering rivers? If by some chance (for example, due to bank collapse) a small bend is formed in the river, then the water will have to flow around it in a curved path, and the surface on the concave side of the bend will be slightly higher than on the convex side, creating a pressure gradient to maintain the rotational motion of the fluid along the curved path. Friction against the bottom of the river bed will again lead to secondary circulation (imagine half of Fig.\ref {fig1}). This secondary flow carries away eroded material from the concave side and deposits it as mud on the convex side of the bend. As a result, the convex side of the bend will be shallow and muddy, and the concave side will be deep, eroding the adjacent bank, and the bend will become more pronounced over time. Moreover, as Einstein notes, since the circular motion has inertia, the secondary circulation, and hence the asymmetry of erosion, will reach its maximum beyond the point of greatest curvature, and as a result, over time, the bend will shift in the direction of the flow.

A few remarks are perhaps in order here. Einstein is notorious for not reading the scientific literature of his day\footnote{We have his own confession to Ehrenfest: "My dear friend, do you think that I am in the habit of reading papers written by others?" \cite{Rumer_2001}.} Einstein was apparently unaware that the teacup experiment had been discussed by James Thomson in 1857 as an analogy to atmospheric circulation and in 1876 in connection with the sinuosity of a river \cite{Alpher_1960,Goldstein_1952}. Moreover, both the secondary circulation in the stirred teacup and the reason for the sinuosity of the river are very clearly explained in Edwin Edser's 1911 textbook of general physics \cite{Edser_1911}.

In reality, river meandering is a much more complex phenomenon, the study of which has generated an extensive literature \cite{Callander_1978,Kleinhans_2024}. Einstein's contributions are rarely cited in this extensive literature \footnote{Even his own son, a renowned professor of hydraulic engineering, was apparently not familiar with Einstein's explanation of meandering and makes no mention of it \cite{Einstein-Shen_1964}.}.

Variations of the teacup experiment can produce surprising results. For example, in a flow caused by the rotation of the bottom plate of a partially filled stationary cylindrical vessel, the shape of the free surface of the liquid can break axial symmetry and form a rigidly rotating polygon with a different rotational velocity than that of the plate when the rotational velocity of the plate becomes sufficiently large \cite{Vatistas_1990,Jansson_2006}.

Regarding the tea leaf effect, I was intrigued by one remark: ``It is said that, with this explanation, Einstein appeased Mrs. Sch\"{o}dinger's curiosity, which he husband could not satisfy'' \cite{Moisy_2003}. The following letter from Schr\"{o}dinger to Einstein, dated April 23, 1926, explains what actually happened: ``I very much enjoyed your delightful explanation of the formation of meanders. It just happens that my wife had asked me about the <<teacup phenomenon>> a few days earlier, but I did not know a rational explanation. She says that she will never stir her tea again without thinking of you'' \cite{Buchwald_2018}.

\section{Rayleigh–B\'{e}nard-Marangoni convection}
Rotation is not the only factor that drives convection. Convection driven by non-uniform heating is an even more common type of fluid motion in the Universe. Rayleigh–B\'{e}nard-Marangoni convection, ``the granddaddy of canonical examples used to study pattern formation and behavior in spatially extended systems'' \cite{Newell_1993}, is perhaps the simplest example of this type of convection.

\begin{figure}[h]
    \centering
    \includegraphics[scale=1.0]{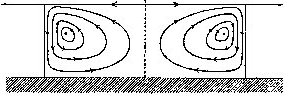}
    \caption{B\'{e}nard's illustration of fluid circulation in a cross-section of a hexagonal cell. }
    \label{fig2}
\end{figure}

Rayleigh–B\'{e}nard-Marangoni convection occurs in a thin layer of fluid heated from below (or cooled from above) \cite{Chandrasekhar_1981,Getling_1998}. The physical mechanism for this phenomenon is as follows \cite{Chandrasekhar_1981,Velarde_1980}. Due to thermal expansion, the fluid at the bottom will be lighter than the fluid at the top, and the fluid will tend to redistribute itself to correct this gravitationally unstable arrangement. However, the dissipative effects of fluid viscosity and heat diffusion prevent such a redistribution, and for this reason the temperature difference must exceed a certain critical value before instability can occur. When the critical gradient is exceeded, the warm fluid tends to rise everywhere, and the cold fluid tends to sink everywhere. However, both phenomena cannot occur simultaneously and in the same place, and this dilemma is resolved by the spontaneous division of the layer into a mosaic of convective cells, in each of which the liquid circulates along a closed path.
In 1900, Henri B\'{e}nard experimentally observed the transition of a thin layer of spermaceti (whale blubber), heated from below, from a spatially homogeneous stationary state to an array of convection cells constituting a periodic mosaic in space. With careful experimental planning, this mosaic consisted of stationary patterns of nearly identical hexagons arranged like a honeycomb. B\'{t}nard's drawing of fluid circulation in a hexagonal cell vortex is shown in Fig.\ref{fig2} \cite{Fauve_2017}. The hot fluid rises along the central axis of the cell, cools after circulating along the upper free surface, sinks down along the edges of the hexagonal cell, and heats up after moving along the heated lower plate. The analogy with secondary circulation in a cup of tea is obvious.

A theoretical explanation for B\'{e}nard's experimental results was given in 1916 by Lord Rayleigh \cite{Rayleigh_1916} using the physical mechanism described above. In new experiments, B\'{e}nard attempted to test Rayleigh's quantitative predictions. The wavelength of the observed periodic pattern in the experiments agreed well with theory, but a strong discrepancy was observed for the convection threshold. The reason for this discrepancy was not fully elucidated until 1956 \cite{Block_1956,Fauve_2017}. It turned out that buoyancy was not responsible for the convection in B\'{e}nard's experiments, as Rayleigh had assumed. Instead, a second instability mechanism was discovered, involving surface tension gradients, which usually coexists with Rayleigh's buoyancy mechanism but dominates in thin layers. This instability mechanism, now called B\'{e}nard-Marangoni convection, operates in the presence of a free fluid surface that can move and deform. Pure Rayleigh convection, driven by buoyancy, can arise if a thin layer of fluid is confined between two flat horizontal plates and completely fills the space between them. In this case, the basic unit of the convective mosaic is not a hexagon but a long tubular ``roll'', and the repeating unit in the convective pattern consists of two counter-rotating rolls \cite{Velarde_1980}.

The granules observed on the solar surface are convective cells in the solar photosphere and resemble B\'{e}nard cells. However, despite the superficial similarity, solar convection differs significantly and in many respects from Rayleigh–B\'{e}nard–Marangoni convection \cite{Schumacher_2020}. Any conclusions that can be transferred from controlled laboratory experiments on convective instability to some extreme astrophysical or cosmological situation are necessarily qualitative and may turn out to be incorrect. With this caveat, we proceed and blithely draw a conclusion of truly cosmic proportions from Einstein's remark on secondary circulation.

\section{Heavy seeds formation in ekpyrotic universe}
Traditionally, physics is an empirical science, often guided by theoretical insights. Its development so far has confirmed two of T.D. Lee's laws: ``without experimentalists, theorists tend to drift'' and ``without theorists, experimentalists tend to falter'' \cite{Lee_1986}. However, with cosmology on the one hand, where ``some models have their roots in very speculative physics (e.g., superstring theory), and others are simply {\it ad hoc}'' \cite{Turner:2018},  and string theory on the other, where empirical testing does not seem feasible in a reasonable future, we are faced with a certain paradigm shift in science, according to which the ``criteria  of  rationality  are  inherently social:  values  only  have  force  if  they  are  shared'' \cite{vanDongen_2021}.

I am not saying that the state of affairs in high energy theory/string theory/etc. is such that it is fundamentally divorced from reality. The creation of the Standard Model and the detection of gravitational waves are remarkable achievements. However, it is true that many popular recent ideas ``would require happenings every bit as miraculous as the views of religious fundamentalist" {\cite{Hoyle_1982}}.

One interesting and widely accepted speculative idea in cosmology is that our universe is a thin shell (brane) in a higher-dimensional space \cite{Rubakov_1983,Gogberashvili_1998,Randall_1999}, and that the universe as we know it arose from the collision of two branes \cite{Dvali_1998,Gabadadze_1999,Khoury_2001}. The theoretical basis for such models is the brane world of superstring theory \cite{Kakushadze_1998,Pease_2001,Antoniadis_1998}.

\begin{figure}[h]
    \centering
    \includegraphics[scale=0.4]{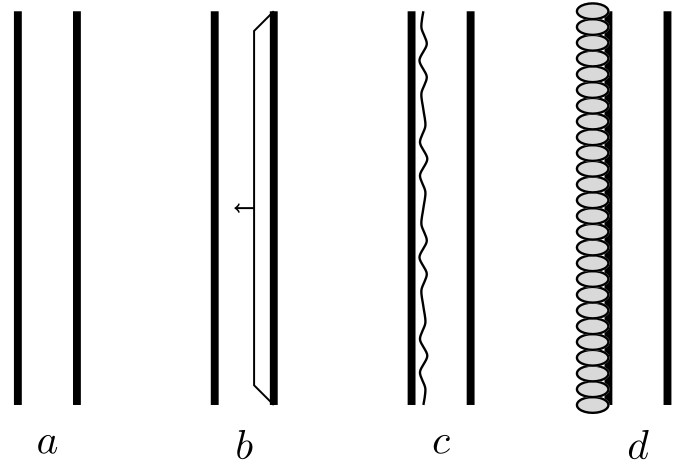}
    \caption{Ekpyrotic universe scenario with final Rayleigh–B\'{e}nard-Marangoni type convection.}
    \label{fig3}
\end{figure}

The idea of one such model, the so-called ekpyrotic universe scenario \cite{Khoury_2001}, is shown schematically in Fig. \ref{fig3}. The initial state of such a universe consists of two static massive superstring branes and a curved geometry in the intermediate volume (a). Spontaneously, the bulk brane separates from the hidden brane in some region of space, the edges of which expand outward at the speed of light, while the interior moves very slowly toward the opposite visible brane (b). Quantum fluctuations create ripples as the light bulk brane moves, and when the bulk brane collides with the visible brane, the ripples cause the different regions to collide at slightly different times, thereby creating density fluctuations in the visible universe (c).

Here we note an analogy with the conditions for the occurrence of Rayleigh–B\'{e}nard-Marangoni type convection. The collision is thought to fill the visible brane with matter in the form of an ideal fluid \cite{Khoury_2001}.  This fluid is very thin in the extra bulk dimension (we assume codimension one braneworld models in an effective 5-dimensional spacetime {\cite{Lehners:2008vx}}. Although codimension two braneworld models in an effective 6-dimensional spacetime have also been considered, in such models the behavior of gravity is qualitatively very different {\cite{Maartens:2010ar}}, which would make the analogy with convection much weaker). Moreover, an ideal fluid (excluding superconductivity) is just an abstraction, and from the point of view of hydrodynamics, the inclusion of the concepts of viscosity in the macroscopic theory of cosmic fluid would seem more natural \cite{Brevik_2017}. Effective gravity is provided by a mechanism that ensures that ordinary matter is attached to the visible brane, although it is unclear which effects dominate: buoyancy or surface tension. Finally, extreme and uneven heating on one side is provided by the colliding bulk brane. Therefore, at first glance, there is every reason for the emergence of Rayleigh–B\'{e}nard–Marangoni type convection, leading to more significant density inhomogeneities than quantum fluctuations (d). 

Such larger-than-usually-assumed density perturbations could perhaps help explain the existence of mysterious magnetic fields that are uniform over astronomically large scales \cite{Berezhiani_2003}. The putative Rayleigh–B\'{e}nard–Marangoni type convection may also be responsible for creating the observed regular or quasi-regular structures in galaxy distributions on large scales \cite{Einasto_1997,Ryabinkov_2023,Einasto_2013}.

We admit that the connection with the Rayleigh–B\'{e}nard–Marangoni convection is highly speculative and, at first glance, probably far-fetched. Even if the proposed convection is indeed viable, it would take place in a four-dimensional space with three-dimensional free hypersurfaces and may differ significantly from ordinary three-dimensional convection, where the free surfaces are two-dimensional. Note that numerical simulations revealed both similarities and significant differences between two- and three-dimensional Rayleigh–B\'{e}nard convection {\cite{Lohse_2013}}.

On the other hand, of course, it's interesting to talk about colliding branes, inflation, etc., but is there any compelling experimental or observational evidence that we could even dream of?

We can point out at least two theoretical predictions of the ekpyrotic model that can be tested observationally. The model, like inflation, can predict a nearly scale-invariant spectrum of scalar perturbations that act as seeds of the large-scale structure of the Universe. The spectrum of scalar perturbations can be inferred from measurements of the temperature anisotropy of the cosmic microwave background, and observations already constrain ekpyrotic models, which typically predict a noticeable non-Gaussianity of the spectrum {\cite{Planck:2019kim}}.

Tensor perturbations (the primordial gravitational wave background) are also generated and can potentially be observed, although the canonical inflationary and ekpyrotic models give radically different predictions.  {\cite{Lehners:2008vx}}.  Recently, several pulsar timing array  collaborations have reported the presence of an isotropic stochastic background of gravitational waves with a blue spectrum which might be
a potential hint for the Ekpyrotic bouncing cosmology {\cite{Qiu:2024sdd}}.

\section{Concluding remarks}
As observations become more precise, it becomes clear that the $\Lambda$CDM cosmological model has some problems, and its proponents replace one fine-tuning with another to fix them. However, one could argue that constantly replacing $\Lambda$CDM problems with other problems, while intellectually fun, simply moves the problem from one side of the table to the other. A natural question arises: what is the benefit of this activity then?

To argue that it is useful to have a large stock of ideas, even exotic ones, we can refer to the time-tested strategy of the living cell, a strategy that makes life vibrant and resilient. As long as the surrounding conditions are stable, the properties of the organism, its phenotype, do not change, although genetic mutations occur all the time and accumulate in the genome. The cell simply has a mechanism that silences the wrong genes. 

But when a crisis situation occurs, for example, an asteroid falls to the earth, and the surrounding conditions change abruptly, the molecular mechanisms of the cell open the floodgates, and also sharply increase the frequency of mutations, and the accumulated genetic variations are presented to the judgment of reality. Of course, most of them are doomed to failure, but there will be those that are suitable for the new conditions, and ``gray, worthy friend, is all your theory
and green the golden tree of life" {\cite{Goethe_1856}}.

The main idea of this note is that convection may have played a significant role in the early Universe, in particular in creating the initial density inhomogeneities. We present only a qualitative discussion, much in the spirit of Einstein's paper \cite{Einstein_1926}, within the currently popular ekpyrotic universe model.

Of course, at least a toy model calculation that would make it easier to understand whether this idea is viable at all would be desirable. However, even formulating such a model that would be more or less realistic is not an easy task. We are not aware of any discussion of Rayleigh–B\'{e}nard–Marangoni convection in the context of general relativity, let alone in models with extra dimensions, and there is probably a good reason for this.

The Rayleigh–B\'{e}nard–Marangoni convection theories are based on the Navier–Stokes equations (in the Boussinesq approximation) {\cite{Chandrasekhar_1981}}, which are already complex equations. But we need their relativistic generalizations.

It may appear that relativistic generalization is straightforward, essentially a matter of imposing the original Navier–Stokes equations in the local reference frame of the fluid at rest. Indeed two relativistic versions of hydrodynamics were formulated by Eckart (1940) and by Landau and Lifshitz (1959) long ago {\cite{Andersson:2020phh}}. These models have been and are still used in many applications. However, upon closer examination, it turns out that they suffer from serious pathologies, namely, they are unstable, non-causal, and do not have a well posed initial value formulation. Therefore, at the very least, they should be used with caution {\cite{Andersson:2020phh}}.

The reason for the pathologies is simple: the equations of the supposedly relativistic models form a parabolic system, like the original Navier–Stokes equations, and thus transmit signals at arbitrarily high speeds {\cite{Geroch_1995}}.

Of course, hyperbolic theories of relativistic hydrodynamics can be imagined, and examples, from a very large number of possibilities, include theories developed by Stewart (1977), Israel and Stewart (1979), and Carter (1991) {\cite{Andersson:2020phh}}.

Hyperbolic (causal) relativistic generalizations of the Navier-Stokes equations significantly increase the number of dynamical fields required to describe a fluid {\cite{Geroch_1995,Lindblom_1996}}. Can we develop a "correct", empirically sound relativistic theory of cosmic fluids in the early Universe from a variety of possibilities, if these new extra degrees of freedom due to additional dynamical fields have never been directly observed in real laboratory fluids? The answer is probably no: we should not expect to be able to use observations to choose a preferred theoretical description, since usually the different models become empirically indistinguishable on the very short time scales of microscopic particle interactions, and the differences between these theories become important only in regimes in which all such theories are expected to break down {\cite{Geroch_1995,Lindblom_1996,Andersson:2020phh}}.

Sometimes physics is just fun. The idea that the same mechanism that causes tea leaves to gather in the center of a stirred cup, or that creates meandering rivers, is at work in the early universe, helping to form the seeds of supermassive black holes, may be a crazy idea. But is it crazy enough to be true?

\section*{Acknowledgments}
The author would like to thank the anonymous reviewer for his/her open-minded and impartial review, which allowed the author to clarify some aspects of the manuscript more vividly.

\bibliography{SMBH_seeds}

\begin{thebibliography}{10}

\bibitem{Dolgov_1993}
A.~Dolgov and J.~Silk, {\em Phys. Rev. D} {\bf 47}, 4244  (1993).

\bibitem{Dodelson_2021}
S.~Dodelson and F.~Schmidt, {\em Modern cosmology}, 2 edn. (Academic Press, London, 2021).

\bibitem{Turner:2018}
M.~S. Turner, {\em Found. Phys.} {\bf 48}, 1261  (2018), \href{http://arxiv.org/abs/2109.01760}{{\ttfamily arXiv:2109.01760 [astro-ph.CO]}}.

\bibitem{Perivolaropoulos:2021}
L.~Perivolaropoulos and F.~Skara, {\em New Astron. Rev.} {\bf 95},   101659  (2022), \href{http://arxiv.org/abs/2105.05208}{{\ttfamily arXiv:2105.05208 [astro-ph.CO]}}.

\bibitem{Dolgov_2018}
A.~D. Dolgov, {\em Physics-Uspekhi} {\bf 61},   112  (2018).

\bibitem{Silk_2024}
J.~Silk, M.~C. Begelman, C.~Norman, A.~Nusser and R.~F.~G. Wyse, {\em Astrophys. J. Lett.} {\bf 961},   L39  (2024).

\bibitem{Woods_2019}
T.~E. Woods, B.~Agarwal, V.~Bromm, A.~Bunker, K.-J. Chen, S.~Chon, A.~Ferrara, S.~C.~O. Glover, L.~Haemmerlé, Z.~Haiman and et~al., {\em Publ. Astron. Soc. Australia} {\bf 36},   e027  (2019).

\bibitem{Volonteri_2010}
M.~Volonteri, {\em Astron. Astrophys. Rev.} {\bf 18}, 279  (2010).

\bibitem{Carr:2003}
B.~J. Carr, {\em Lect. Notes Phys.} {\bf 631}, 301  (2003), \href{http://arxiv.org/abs/astro-ph/0310838}{{\ttfamily arXiv:astro-ph/0310838}}.

\bibitem{Carr:2023}
B.~Carr, S.~Clesse, J.~Garcia-Bellido, M.~Hawkins and F.~Kuhnel, {\em Phys. Rept.} {\bf 1054}, 1  (2024), \href{http://arxiv.org/abs/2306.03903}{{\ttfamily arXiv:2306.03903 [astro-ph.CO]}}.

\bibitem{Khlopov:2004}
M.~Y. Khlopov, S.~G. Rubin and A.~S. Sakharov, {\em Astropart. Phys.} {\bf 23}, 265  (2005), \href{http://arxiv.org/abs/astro-ph/0401532}{{\ttfamily arXiv:astro-ph/0401532}}.

\bibitem{Padmanabhan_2023}
H.~Padmanabhan and A.~Loeb, {\em Astrophys. J. Lett.} {\bf 953},  ~L4  (2023).

\bibitem{Tkachev:2023}
M.~V. Tkachev, S.~V. Pilipenko, E.~V. Mikheeva and V.~N. Lukash, {\em Mon. Not. Roy. Astron. Soc.} {\bf 527}, 1381  (2023), \href{http://arxiv.org/abs/2307.13774}{{\ttfamily arXiv:2307.13774 [astro-ph.CO]}}.

\bibitem{Einstein_1926}
A.~Einstein, {\em Naturwissenschaften} {\bf 14}, 223  (1926).

\bibitem{Yogananda_2000}
C.~S. Yogananda and A.~Einstein, {\em Resonance} {\bf 5}, 105  (2000).

\bibitem{Rumer_2001}
Y.~B. Rumer, {\em Physics-Uspekhi} {\bf 44}, 1082  (2001).

\bibitem{Alpher_1960}
R.~A. Alpher and R.~Herman, {\em Am. J. Phys.} {\bf 28}, 748  (1960).

\bibitem{Goldstein_1952}
S.~Goldstein (ed.), {\em Modern Developments in Fluid Dynamics Vol. 1} (Clarendon Press, Oxford, 1952).

\bibitem{Edser_1911}
E.~Edser, {\em General physics for students: a textbook of the fundamental properties of matter} (Macmillan, London, 1911).

\bibitem{Callander_1978}
R.~A. Callander, {\em Annu. Rev. Fluid Mech.} {\bf 10}, 129  (1978).

\bibitem{Kleinhans_2024}
M.~G. Kleinhans, W.~J. McMahon and N.~S. Davies, {\em Geol. Soc. Spec. Publ} {\bf 540}, SP540  (2024).

\bibitem{Einstein-Shen_1964}
H.~A. Einstein and H.~W. Shen, {\em J. Geophys. Res.} {\bf 69}, 5239  (1964).

\bibitem{Vatistas_1990}
G.~H. Vatistas, {\em J. Fluid Mech.} {\bf 217}, 241–  (1990).

\bibitem{Jansson_2006}
T.~R.~N. Jansson, M.~P. Haspang, K.~H. Jensen, P.~Hersen and T.~Bohr, {\em Phys. Rev. Lett.} {\bf 96},   174502  (2006).

\bibitem{Moisy_2003}
F.~Moisy, T.~Pasutto, G.~Gauthier, P.~Gondret and M.~Rabaud, {\em Europhysics News} {\bf 34}, 104  (2003).

\bibitem{Buchwald_2018}
D.~K. Buchwald, J.~Illy, J.~N. James, A.~M. Hentschel, M.~J. Teague, A.~Aebi and K.~Hentschel (eds.), {\em The {Berlin} Years: Writings and Correspondence, {June 1925--May 1927}}, The collected papers of Albert Einstein, Vol.~15 (Princeton University Press, Princeton, 2018).

\bibitem{Newell_1993}
A.~C. Newell, T.~Passot and J.~Lega, {\em Annu. Rev. Fluid Mech.} {\bf 25}, 399  (1993).

\bibitem{Chandrasekhar_1981}
S.~Chandrasekhar, {\em {H}ydrodynamic and hydromagnetic stability} (Dover Publications, New York, 1981).

\bibitem{Getling_1998}
A.~Getling, {\em Rayleigh-B\'{e}nard Convection: Structures and Dynamics} (World Scientific, Singapore, 1998).

\bibitem{Velarde_1980}
M.~G. Velarde and C.~Normand, {\em Sci. Am.} {\bf 243}, 92  (1980).

\bibitem{Fauve_2017}
S.~{Fauve}, {\em Comptes Rendus Physique} {\bf 18}, 531  (2017).

\bibitem{Rayleigh_1916}
L.~Rayleigh, {\em Phil. Mag. Ser. 6} {\bf 32}, 529  (1916).

\bibitem{Block_1956}
M.~J. Block, {\em Nature} {\bf 178}, 650  (1956).

\bibitem{Schumacher_2020}
J.~Schumacher and K.~R. Sreenivasan, {\em Rev. Mod. Phys.} {\bf 92},   041001  (2020).

\bibitem{Lee_1986}
T.~D. Lee, {\it {The evolution of weak interactions}} CERN-86-07 (9, 1986).

\bibitem{vanDongen_2021}
J.~van Dongen, {\em Stud. Hist. Phil. Sci. A} {\bf 89}, 164  (2021), \href{http://arxiv.org/abs/2105.14342}{{\ttfamily arXiv:2105.14342 [physics.hist-ph]}}.

\bibitem{Hoyle_1982}
F.~Hoyle, {\em Annu. Rev. Astron. Astrophys.} {\bf 20}, 1  (1982).

\bibitem{Rubakov_1983}
V.~Rubakov and M.~Shaposhnikov, {\em Phys. Lett. B} {\bf 125}, 136  (1983).

\bibitem{Gogberashvili_1998}
M.~Gogberashvili, {\em EPL} {\bf 49}, 396  (2000), \href{http://arxiv.org/abs/hep-ph/9812365}{{\ttfamily arXiv:hep-ph/9812365}}.

\bibitem{Randall_1999}
L.~Randall and R.~Sundrum, {\em Phys. Rev. Lett.} {\bf 83}, 4690  (1999), \href{http://arxiv.org/abs/hep-th/9906064}{{\ttfamily arXiv:hep-th/9906064}}.

\bibitem{Dvali_1998}
G.~R. Dvali and S.~H.~H. Tye, {\em Phys. Lett. B} {\bf 450}, 72  (1999), \href{http://arxiv.org/abs/hep-ph/9812483}{{\ttfamily arXiv:hep-ph/9812483}}.

\bibitem{Gabadadze_1999}
G.~R. Dvali and G.~Gabadadze, {\em Phys. Lett. B} {\bf 460}, 47  (1999), \href{http://arxiv.org/abs/hep-ph/9904221}{{\ttfamily arXiv:hep-ph/9904221}}.

\bibitem{Khoury_2001}
J.~Khoury, B.~A. Ovrut, P.~J. Steinhardt and N.~Turok, {\em Phys. Rev. D} {\bf 64},   123522  (2001), \href{http://arxiv.org/abs/hep-th/0103239}{{\ttfamily arXiv:hep-th/0103239}}.

\bibitem{Kakushadze_1998}
Z.~Kakushadze and S.~H.~H. Tye, {\em Nucl. Phys. B} {\bf 548}, 180  (1999), \href{http://arxiv.org/abs/hep-th/9809147}{{\ttfamily arXiv:hep-th/9809147}}.

\bibitem{Pease_2001}
R.~Pease, {\em Nature} {\bf 411}, 986  (2001).

\bibitem{Antoniadis_1998}
I.~Antoniadis, N.~Arkani-Hamed, S.~Dimopoulos and G.~Dvali, {\em Physics Letters B} {\bf 436}, 257  (1998).

\bibitem{Lehners:2008vx}
J.-L. Lehners, {\em Phys. Rept.} {\bf 465}, 223  (2008), \href{http://arxiv.org/abs/0806.1245}{{\ttfamily arXiv:0806.1245 [astro-ph]}}.

\bibitem{Maartens:2010ar}
R.~Maartens and K.~Koyama, {\em Living Rev. Rel.} {\bf 13},  ~5  (2010), \href{http://arxiv.org/abs/1004.3962}{{\ttfamily arXiv:1004.3962 [hep-th]}}.

\bibitem{Brevik_2017}
I.~Brevik, O.~Gr\o{}n, J.~de~Haro, S.~D. Odintsov and E.~N. Saridakis, {\em Int. J. Mod. Phys. D} {\bf 26},   1730024  (2017), \href{http://arxiv.org/abs/1706.02543}{{\ttfamily arXiv:1706.02543 [gr-qc]}}.

\bibitem{Berezhiani_2003}
Z.~Berezhiani and A.~D. Dolgov, {\em Astropart. Phys.} {\bf 21}, 59  (2004), \href{http://arxiv.org/abs/astro-ph/0305595}{{\ttfamily arXiv:astro-ph/0305595}}.

\bibitem{Einasto_1997}
J.~Einasto, M.~Einasto, S.~Gottl{\"o}ber, V.~M{\"u}ller, V.~Saar, A.~A. Starobinsky, E.~Tago, D.~Tucker, H.~Andernach and P.~Frisch, {\em Nature} {\bf 385}, 139  (1997).

\bibitem{Ryabinkov_2023}
A.~I. Ryabinkov and A.~D. Kaminker, {\em Mon. Not. Roy. Astron. Soc.} {\bf 527}, 1813  (2023).

\bibitem{Einasto_2013}
J.~Einasto, {\em Dark Matter and Cosmic Web Story} (World Scientific, Singapore, 2013).

\bibitem{Lohse_2013}
E.~P. van~der Poel, R.~J. A.~M. Stevens and D.~Lohse, {\em J. Fluid Mech.} {\bf 736},   177–194  (2013).

\bibitem{Planck:2019kim}
Planck Collaboration, Y.~Akrami {\em et~al.}, {\em Astron. Astrophys.} {\bf 641},  ~A9  (2020), \href{http://arxiv.org/abs/1905.05697}{{\ttfamily arXiv:1905.05697 [astro-ph.CO]}}.

\bibitem{Qiu:2024sdd}
T.~Qiu and M.~Zhu, {\em e-print} {\bf arXiv:2408.06582}  (2024), \href{http://arxiv.org/abs/2408.06582}{{\ttfamily arXiv:2408.06582 [gr-qc]}}.

\bibitem{Goethe_1856}
J.~W.~V. Goethe, {\em Faust: a tragedy, translated from the German of Goethe, with notes by Charles T. Brooks} (Ticknor and Fields, Boston, 1856).

\bibitem{Andersson:2020phh}
N.~Andersson and G.~L. Comer, {\em Living Rev. Rel.} {\bf 24},  ~3  (2021), \href{http://arxiv.org/abs/2008.12069}{{\ttfamily arXiv:2008.12069 [gr-qc]}}.

\bibitem{Geroch_1995}
R.~Geroch, {\em J. Math. Phys.} {\bf 36}, 4226  (1995).

\bibitem{Lindblom_1996}
L.~Lindblom, {\em Annals of Physics} {\bf 247}, 1  (1996).

\end{thebibliography}

\end{document}